# Experimental Investigation of Thermal Boundary Layers and Associated Heat Loss for Transient Engine-Relevant Processes Using HRCARS and Phosphor Thermometry

Anthony O. Ojo, David Escofet-Martin, Joshua Collins, Gabriele Falconetti, and Brian Peterson

Institute for Multiscale Thermofluids, School of Engineering, University of Edinburgh, The King's Buildings, Mayfield Road, Edinburgh, EH9 3BF, UK.

**Corresponding author:** Anthony O. Ojo, Email: anthony.ojo@ed.ac.uk

**Abstract:** Design of efficient, downsized piston engines requires a thorough understanding of transient near-wall heat losses. Measurements of the spatially and temporally evolving thermal boundary layer are required to facilitate this knowledge. This work takes advantage of hybrid fs/ps rotational coherent anti-Stokes Raman spectroscopy (HRCARS) to measure single-shot, wall-normal gas temperatures, which provide exclusive access to the thermal boundary layer. Phosphor thermometry is used to measure wall temperature. Measurements are performed in a fixed-volume chamber that operates with a transient pressure rise/decay to simulate engine-relevant compression/expansion events. This simplified environment is conducive for fundamental boundary layer and heat transfer studies associated with engine-relevant processes. The thermal boundary layer development and corresponding heat losses are evaluated within two engine-relevant regimes: (1) an unburned-gas regime comprised of gaseous compression and (2) a burned-gas regime, which includes high-temperature compression and expansion processes. The time-history of important boundary layer quantities such as gas / wall temperatures, boundary layer thickness, wall heat flux, and relative energy lost at the wall are evaluated through these regimes. During the mild unburned-gas compression, $T_{core}$ increases by 30 K and a thermal boundary layer is initiated with thickness $\delta_T \sim 200$ μm. Wall heat fluxes remain below 6 kW/m², but corresponds to ~6% energy loss per ms. In the burned-gas regime, $T_{core}$ resembles adiabatic flame temperatures, while $T_{wall}$ increases by 16 K. A thermal boundary layer rapidly develops as $\delta_T$ increases from 290-730 μm. Energy losses in excess of 25% occur after flame impingement and slowly decay to ~10% at the end of expansion. Measurements also resolve thermal mixing of fresh- and burned gases during expansion, which yield strong temperature reversals in the boundary layer. Findings are compared to canonical environments and demonstrate the transient thermal boundary nature during engine-relevant processes.

## 1.0. Introduction

Heat loss at gas/wall interfaces plays an important role for the design of cleaner internal combustion (IC) and gas-turbine engines [1,2]. For IC engines, there is emphasis towards smaller engines operating with higher power densities. Such concepts offer gains in energy efficiency (i.e. lower $CO_2$) [3], and are ideal as a co- or secondary-powertrain in electric vehicles. However, these downsized, boosted engines are subject to increased transient heat transfer from larger surface-to-volume ratios and more reactions occurring near surfaces. This heat loss yields appreciable thermal stratification into the core-gases [4–8], which directly influences processes such as mixing, heat release, and pollutant formation [2,9,10]. Indeed, the heat exchange at chamber surfaces is a parasitic energy loss, which determines engine performance, efficiency and emissions.

The mass and energy transfer within the velocity and thermal boundary layer largely govern near-wall heat loss. In engines, these boundary layers are unsteady; the outer fluid is not constant in velocity, temperature, or pressure [11]. This unsteady behavior deviates from boundary layer theory [12] and is not fully understood. Most wall functions in heat transfer models do not account for such deviations. Consequently, they struggle to predict engine heat loss accurately. This is troubling for engine simulations employing such models because they become a less effective design tool.

Laser diagnostics have been instrumental to understand unsteady boundary layer behavior. Particle tracking velocimetry has been widely used to investigate velocity boundary layers in engines. Detailed measurements have resolved transient boundary layer events such as vortical structures in the boundary layer [13,14], flow acceleration from an approaching flame front [15,16], and larger scale flow features that yield strong pressure gradients [17,18]. These events govern the energy transport from the boundary layer, and show large deviations from law-of-the-wall predictions. Direct numerical simulations in engine environments also support such findings [11,19]. These studies have contributed to the development of non-equilibrium wall models, which yield improved predictions of near-wall velocities and wall shear stress [18,20].

While the *velocity* boundary layer has been studied in detail, the *thermal* boundary layer has been studied far less. This is primarily due to a limited number of diagnostics that can resolve near-wall gas temperatures with high spatial resolution and accuracy. Laser induced fluorescence (LIF) has been widely used to measure gas temperatures in engines (e.g., [4–10]). Due to challenges with accurate calibration and vignetting near surfaces, many LIF investigations use a normalization scheme, which yields a relative temperature rather than an absolute temperature [5–9]. While LIF studies have demonstrated the magnitude of thermal stratification originating from chamber surfaces [4–9], relative temperature distributions, low signal-to-noise ratios and coarse spatial resolution has limited the ability to resolve the thermal boundary layer using LIF. Nonetheless, such studies have emphasized the need for detailed thermal boundary layer measurements in order to predict the resulting thermal stratification.

Coherent anti-Stokes Raman spectroscopy (CARS) has been the most widely used technique to measure gas temperatures within boundary layers with high accuracy and precision [21–26]. The majority of CARS thermometry applied within boundary layers have employed nanosecond CARS techniques, which provide 0-D point-wise measurements [21–24]. While such data has provided useful information about the average boundary layer structure, such data is comprised of statistically independent and often uncorrelated events. Such data is less conducive to describe transient heat transfer associated with individual events.

Developments in short-pulse femtosecond (fs) and picosecond (ps) CARS techniques have enabled new opportunities for 1D spatially resolved gas-phase temperature measurements [25–32]. Bohlin et al. [25,26] demonstrated the use of hybrid fs/ps rotational CARS (referred to as HRCARS hereafter) to resolve the instantaneous thermal boundary layer within a flame-wall interaction (FWI) burner. Such measurements have impressive spatial resolution (35-61 μm) and precision (~1-3%). HRCARS has further demonstrated its capability for simultaneous temperature, pressure and species concentration measurements near walls [32].

While HRCARS has been demonstrated in near-wall applications, it has only recently been used for engine-related heat transfer studies. Escofet-Martin et al. [27] utilized HRCARS to resolve thermal boundary layers associated with three gas-wall interactions relevant to engines: (1) polytropic compression of unburned gases, (2) FWI, and (3) post-flame and gas expansion. Studies were conducted in a fixed volume chamber (FVC), which operated with a mild compression/expansion process at engine timescales. The authors combined HRCARS with phosphor thermometry and CH* imaging to measure wall temperature and track the flame front progression during FWI. The combined measurements captured transient processes within the boundary layer, which described transient heat fluxes and gas-phase thermal stratification.

In this work, we capitalize on our measurements in [27] to study the transient near-wall heat loss associated with the thermal boundary layer in more detail. This work extends beyond [27] by evaluating the evolution of key boundary layer quantities during the polytropic compression regime and during several distinct regimes of the post-flame and gas expansion events. In addition, we evaluate the highly transient processes during late expansion/exhaust, which reveal temperature reversals and thermal mixing within the outer boundary layer region. Further analyses are performed to describe the thermal boundary shape relative to laminar and turbulent conditions for flow over a flat plate. Experiments provide a database to understand transient heat losses in a simplified, yet non-canonical environment, and are intended to support numerical wall model development.

## 2.0. Experimental setup

### 2.1. Fixed-volume chamber (FVC)

Measurements were performed in an optically accessible FVC. Figure 1 shows the schematic of the FVC. The chamber features test section (150 cm$^3$, surface/volume = 2.32 cm$^{-1}$) and back-pressure section (6 cm$^3$). Separating these sections is an orifice plate of 6 mm thickness with 81 equidistant holes of 0.5 mm diameter. The test section features an inlet port, spark plug, and two pressure transducers. The test section emulates a simplified piston engine geometry at top-dead center, including a crevice region, which can be adjusted for variable crevice spacings (0.5-5 mm). In this work, the crevice spacing is 1.25 mm. Fused silica (FS) windows on the top, side, and front plates provide optical access into the test section. The option exists to replace individual windows with a stainless-steel counterpart. Metal components are comprised of 304 stainless-steel. In the current study, the chamber walls are not externally cooled.

The chamber's operation is modelled after [33] and demonstrated in [27]. The FVC is evacuated to ~ 0.02 bar using a vacuum pump. A homogeneous methane/air mixture ($\Phi$=0.9) is introduced into the FVC until an initial pressure of 1.02 bar is reached. The mixture is ignited via the spark plug. Heat release initiates an exponential pressure rise and compresses the unburned-gas ahead of the flame. At a preselected pressure of 2.02 bar, a dump-valve in the back-pressure region is actuated. The flow is choked through the orifice plate, controlling the exiting mass flow rate and providing an exponential pressure decay.

The average pressure-time curve from 42 experiments is shown in Fig. 2a. The pressure curve, although mild, simulates compression and expansion events with timescales similar to an IC engine. The grey area surrounding the pressure curve represents the standard deviation. The low standard deviation demonstrates that the FVC's operation, specifically the flame propagation and exhaust events, is repeatable. This repeatable operation allows us to develop a coherent database from a variety of diagnostic measurements, which can be performed in separate experiments.

The progression of the flame front, which further describes the chamber operation, is shown in Figs. 2b and c. The flame progression is captured using CH* chemiluminescence imaging. The sequence of flame images presented in Fig. 2 are acquired from a single experiment, where CH* images were recorded through the side and front windows simultaneously (see Fig. 1b). Details of the CH* imaging are described in Sect. 2.4. The flame progression is chronicled with respect to a reference time described in Fig. 2. The reference time, $t = 0$, refers to the time the flame front arrives at $y = 2$ mm from the wall within the HRCARS measurement volume. Although HRCARS measurements were not performed simultaneously for the image sequence presented in Fig. 2, the position of the HRCARS volume is shown in Fig. 2b.

After the flame is initiated at the spark plug, it progresses towards the opposite end of the FVC. The flame reaches the HRCARS volume ~35 ms after ignition and the flame resembles a skewed-parabola shape within the 2D images. At $t \sim 7$ ms, the flame reaches the end of the FVC and enters the crevice. The flame penetrates only a short distance (< 25 mm) into the crevice before the dump-valve is actuated at $t \sim 28$ ms. The flame progression is slowed significantly upon valve activation and reaches a maximum distance as the peak pressure is reached at $t = \sim 36$ ms. Shortly afterwards, the flame quenches within the crevice. As a result, a large volume of unburned gas exists within the crevice region under these operating conditions. These gases leave the crevice region during expansion/exhaust and is discussed further in Sect. 4.3.2.

The use of methane/air mixtures ($\Phi=0.9$) and an initial pressure of 1.02 bar provides a mild compression and expansion process within the FVC. The use of different fuels (e.g. ethylene) and higher initial pressures increase the rate of compression and magnitude of compression. Such operating conditions are the focus of future work. However, a brief description of findings from ethylene/air mixtures at an initial pressure of 2 bar is presented in Sect. 4.3.2 to describe the outgassing from the crevice region during expansion/exhaust.

*2.2. HRCARS*

For gas thermometry, HRCARS was used to provide spatially resolved 1D measurements. The HRCARS setup comprises of a Ti:Sapphire laser (Coherent) which delivers 35 fs pulses with 3 mJ/pulse at 1 kHz (pump/stokes pulse, 800 nm). This laser is locked, through the synchrolock, to the 9[th] harmonic oscillator frequency (80 MHz) of a diode-pumped Nd:YAG (Ekspla), providing 22 ps pulses with 15 mJ/pulse at 50 Hz (probe pulse, 532 nm). The synchrolock provides a small jitter (< 1 ps) between the fs and ps pulses. The probe delay employed for the HRCARS measurements was 76 ps, which is discussed in [27]. By using a pair of dispersion compensation mirrors, the pump/Stokes beam is pre-chirped with pulse duration of 37 fs at the probe volume.

The HRCARS setup is shown in Fig. 3. Using a pair of cylindrical lenses (L1 and L2), the probe and pump/Stokes beams were focused into the chamber through the side windows. The HRCARS measurement volume is located ~ 45 mm ($\Delta x$) from the spark plug (see Fig. 2). Each beam intersected above the center of a mild-steel metal strip (50 mm × 2 mm × 10 mm $\Delta x \times \Delta z \times \Delta y$), installed on the top piston shell as a step. This strip is used to minimize beam steering effects. At the beam intersection, the HRCARS measurement volume was estimated to be 0.055 mm × 1.2 mm × 3 mm ($\Delta x \times \Delta z \times \Delta y$). To avoid light scattered from the wall, a knife-edge is imaged into the HRCARS probe volume. With this, the first spatial measurement position is ~ 50 μm from the wall. The probe and anti-Stokes beams were collimated using a cylindrical lens (L3) and these beams were separated using two angle-tunable short-pass filters (SPF). Two cylindrical lenses (L4, L5) imaged the *y*-axis of the HRCARS probe volume into the spectrometer entrance with a ~1.3 magnification ratio. Another cylindrical lens (L6) focused the anti-Stokes signal into a spectrometer (Princeton Instruments, 1800 gr/mm grating), which provided a 0.28 cm$^{-1}$/pixel resolution and instrument linewidth of 1.2 cm$^{-1}$ (FWHM). An EMCCD camera (Andor), operating at 50 Hz and 1 ms exposure, recorded the anti-Stokes (HRCARS) signal. The CCD sensor was binned on chip 2 × 1 such that each 1D-HRCARS spectra correspond to ~20 μm in the object plane (i.e. pixel resolution). A spatial resolution of 40 μm (25-75% resolution) was determined by imaging the knife-edge through the probe beam. To determine the wall normal distance in relation to the HRCARS spectra, the knife-edge was removed, making it is possible to observe the clipped probe beam imaged directly into the camera.

The HRCARS model used is detailed in [32], and pressure-dependent linewidths are included with self-broadened linewidths for $N_2$ [34] and $O_2 - N_2$ for $O_2$ [35]. Additional information such as non-resonant spectra and pressure dependent linewidths are described in [27].

The temporal resolution of HRCARS is constrained to 20 ms, i.e., the probe laser repetition rate (50 Hz). This yielded ~ 7 1D-HRCARS measurements per experiment. A time-delay between ignition and the probe pulse was varied to provide HRCARS measurements at selected times from ignition. In total, experiments were repeated 42 times. The precise control of the time-delay and the high repeatability of the FVC operation provides a coherent dataset that allows us to record the temporally evolving gas temperatures with at least 1 ms intervals from ignition.

In this work, 1D-HRCARS gas temperature measurements are presented for signal-to-noise ratios (SNR) ≥ 15. For instances when SNR < 15, neighboring spectra are binned together to achieve SNR ≥ 15. This was performed for $\Delta y$ up to 0.2 mm. If the SNR remained below 15 after binning, the HRCARS spectra was not analyzed.

*2.3. Phosphor thermometry*

Phosphor thermometry was used to measure wall temperature. As reported in [27], phosphor measurements were originally performed simultaneously with HRCARS and CH* imaging. This is depicted in Fig. 3. In this work, we performed additional phosphor measurements to improve measurements in the burned-gas regime when product gases attenuate the laser fluence. These new measurements were not performed simultaneously with HRCARS, but were performed alongside CH* imaging. As discussed below, good agreement was obtained between [27] and measurements performed here.

The thermographic phosphor (TGP) setup is similar to [27] and shown in Fig. 3. Wall temperatures were measured using the phosphor $Gd_3Ga_5O_{12}$:Cr,Ce (GGG). A Nd:YAG laser (InnoSlab Edgewave, 266 nm, ~10 ns), operating at 1 kHz, was used for excitation. The laser pulse energy was 0.02 mJ, which was reduced from the 0.12 mJ used in [27]. At this lower energy, the measured wall temperature was less sensitive to changes in laser fluence. This is important because product gases present within the burned-gas regime attenuate the laser fluence. At the higher laser energy, the laser attenuation negatively affected wall temperature measurements due to a stronger cross-dependence between laser fluence and the measured temperature [36]. This effect was mitigated using the lower pulse energy.

A mixture of GGG and a HPC binder were deposited as a 2 mm × 10 mm × 0.015 mm ($\Delta x \times \Delta z \times \Delta y$) coating onto the metal strip. In this work, the edge of the coating was displaced 2 mm ($\Delta x$) from the HRCARS probe volume location, which is larger than the 0.2 mm displacement in [27]. This increased displacement provided better optical access for improved signal collection associated with the lower laser energy. Using a pinhole, the excitation beam illuminated an area of ~ 8 mm$^2$ on the phosphor coating. A PMT, fitted with a 725±25 nm bandpass filter, detected the temperature-dependent luminescence decay and a Tektronix (MSO3054) oscilloscope acquired the detected signals. Luminescence decays were processed on a single-shot basis to evaluate individual luminescence lifetimes. The lifetimes were calibrated with temperature by monitoring the temperature of a heated TGP-coated aluminium bar, equipped with a thermocouple and thermal insulation, which cooled to ambient.

As mentioned, gas temperatures from HRCARS were measured 45 mm from the spark plug, while wall temperatures from TGP were measured 47 mm from the spark plug (i.e. 2 mm offset). CH* images reveal the flame progresses at a speed of ~1 m/s between the location of 45 mm and 47 mm from the spark plug. This yields a 2 ms delay of the flame arrive time between these distances. To yield similar flame arrival times between HRCARS and TGP, a time difference of 2 ms was subtracted from the TGP thermometry dataset. Comparison of the measured wall temperature at the HRCARS position performed in [27] and the time-adjusted measurements performed in this work showed remarkable agreement. During FWI, the peak wall temperature change ($\Delta T_{wall(max)}$) is 16 K, compared to the 14 K reported in [27]. This difference is within the experimental uncertainty.

*2.4. CH* Imaging*

A high-speed camera (VEO 710L, Phantom) equipped with a 433±14nm bandpass filter and operating with an exposure time of 500 μs at 2 kHz was used to track the flame front (CH*). The acquired CH* images were processed to extract flame front positions at corresponding reference times. As described in Sect 2.1, the reference time, $t = 0$ is defined as the time the flame front reaches the HRCARS measurement volume. CH* images were calibrated pixel-wise by imaging a target with 1 mm equidistant lines placed inside the FVC.

*2.5 Measurement Uncertainty*

At ambient conditions, the HRCARS measurement uncertainty is $\sigma/T = 0.9\%$. HRCARS measurements were performed in a McKenna burner to assess the uncertainty at high temperatures. McKenna burner measurements were benchmarked against a nanosecond vibrational CARS database [37]. Deviations of our HRCARS from the vibrational CARS measurements were less than 6% (i.e. accuracy). The precision uncertainty of our HRCARS at temperatures above 1200 K is $\sigma/T = 3\%$.

Temperature measurements with TGP were compared to thermocouple readings to assess the uncertainty. Assessment of the phosphor thermometry measurement uncertainty was carried out during temperature calibration. During calibration, measurements were conducted on a TGP-coated aluminium bar, equipped with a thermocouple and thermal insulation. The aluminium bar was heated to 150 K in an oven and was removed to cool to ambient. TGP measurements revealed an accuracy better than 1 K and single shot precision uncertainty ~ 0.35% at 296 K, and ~ 1.3% at 320 K. Measurements were performed on the TGP-coated aluminium bar to evaluate the effect of laser fluence on wall temperature measurements. By using neutral density filters, for a 50% drop in laser energy (from 0.02 mJ/pulse to 0.01 mJ/pulse), the change in measured temperature was estimated as < 1 K which is within the measurement uncertainty of our measurements.

Several quantities are derived from HRCARS and TGP measurements to characterize the thermal boundary development and corresponding heat losses. These quantities along with their associated uncertainty are introduced in Sect. 4.1.

## 3.0. HRCARS for thermal boundary layer measurements

1D HRCARS is a relatively new diagnostic and it provides a powerful opportunity to resolve single-shot temperature profiles normal to surfaces [25,26]. However, 1D HRCARS has only begun to be exploited for boundary layer research [27]. As such, it is important to evaluate its performance to study thermal boundary layer development in transient environments.

This section describes the merits and limitations of 1D HRCARS for thermal boundary layer studies. This evaluation is performed for three prominent temperature regimes discussed in this paper: (1) unburned-gas, (2) flame-wall interaction (FWI) and (3) burned-gas. These regimes consider the relative position of the flame front with respect to the HRCARS measurement volume. Each regime is defined by the reference time, *t*, and is shown in the pressure curve in Fig. 2a.

*3.1. Unburned-gas regime*

The unburned-gas regime considers the reference time *t* < - 2 ms, where the flame approaches, but has not reached the HRCARS volume. In this regime, the unburned gases ahead of the flame are compressed in a polytropic fashion due to gas expansion behind the flame [27,33]. The polytropic compression is the driving force that increases the unburned core-gas temperature. As the compression is mild (1.02 to 1.41 bar), the change in core-gas temperature ($\Delta T_{core} = T_{core,t} - T_{core,initial}$) is also mild (0 – 30 K). The precision of the HRCARS during the unburned-gas regime is 2.7 K ($\sigma/T_{gas} = 0.9\%$.). Consequently, this precision can be a significant percentage $\Delta T_{core}$, and the measurement uncertainty is more noticeable when temperature gradients are small.

Figure 4a shows two example 1D temperature profiles at selected times during the unburned-gas regime. At *t* = -16.8 ms, a weak compression has only yielded a small $\Delta T_{core}$ (~ 6 K). While a boundary layer profile is evident, the precision uncertainty is ~26% of $\Delta T_{core}$ and the data scatter is appreciable.

At $t$ = -3.2 ms, compression is more significant, yielding a larger $\Delta T_{core}$. The temperature profile resembles a distinguishable thermal boundary layer profile with less scatter. In both temperature profiles shown in Fig. 4a, the measurement scatter increases beyond $y$ = 2 mm. In our setup, the knife-edge and wall truncates a portion of the pump/stokes and probe beams. Consequently, the probe pulse contains the highest pulse energy within $y$ = 0-2 mm from the wall. The lower probe pulse energy within $y$ = 2-3 mm is responsible for the larger scatter in this region.

*3.2. FWI regime*

The FWI regime is defined by reference times -2 < $t$ < 9 ms during which the flame sweeps through the HRCARS probe volume. During the FWI regime, the core-gas temperatures increase substantially as described in [27]. The change of the core-gas temperature, $\Delta T_{core}$, is approximately two orders of magnitude larger than the unburned-gas regime.

During FWI, beam steering and high temperatures lead to HRCARS signal loss, such that the temperature through the entire measurement length (0-3 mm) cannot be resolved. Figure 4b shows an example temperature profile during FWI at $t$ = 0.1 ms. Temperature measurements are resolved from the wall until the approximate location of the flame front, which is further evidenced within CH* images (not shown). Beyond the flame front, the SNR < 15, and beyond a certain location, no CARS signal is registered. Low SNR values are partially due to high temperatures, but largely due to beam steering when the flame is aligned tangentially with the HRCARS volume. The complete loss of signal is a result of beam steering.

Escofet-Martin et al. [27], discusses the evolution of temperature profiles during FWI in detail. In this work, we do not evaluate the thermal boundary layer during the FWI regime. This is primarily because we are unable to consistently resolve temperatures beyond the flame, but also because the flame completely modifies the pre-existing thermal boundary layer. We report the HRCARS challenges during the FWI regime because these challenges persist into the early stages of the burned-gas regime.

*3.3. Burned-gas (post-FWI) regime*

The burned-gas regime is defined by reference times 8 < $t$ < 100 ms after the flame has progressed through the HRCARS volume. The burned-gas regime includes the remaining compression and exhaust events. In Sect. 4.1.2, the burned-gas regime is further categorized into three sub-regimes. In this section, we focus on HRCARS measurements from 8 < $t$ < 63 ms, which is classified as the post-FWI regime.

Within the post-FWI regime, gas temperatures are the highest recorded, often exceeding 2,000 K in the core-gas. The maximum temperature measured by our HRCARS system is ~ 2,300 K between 10 ms and 20 ms in the post FWI regime. Figure 4b shows two example temperature profiles during the post-FWI regime. At $t$ = 13.4 ms, shortly after FWI, the effects of beam steering and low HRCARS signal prevent measurements beyond $y$ = 1.1 mm. The 1D HRCARS spectra at $t$ = 13.4 ms is shown in Fig. 4c. The HRCARS signal is stronger near the wall where colder gas temperatures exist. Beyond the white line at $y$ = 1.1 mm, SNR < 15 for which measurements are not reported. As time elapses, SNR increases as the gases cool and more of the measurement length is resolved. The temperature profile at $t$ = 60.4 ms reveals this trend where the gas temperatures are nearly resolved within the entire 3 mm domain. Figure 4d shows the 1D HRCARS spectra at $t$ = 60.4 ms, revealing the higher signals away from the wall. Beyond $t$ > 63 ms, measurements are resolved within the entire 3 mm probe length.

Although it is not possible to resolve gas temperatures within the entire probe length during the post-FWI regime, the resolved temperature profiles often show temperatures plateau away from the wall. Starting at $t$ = 13.4 ms, gas temperatures away from the wall plateau to a value near the adiabatic flame temperature (2,134 K [37]). This plateau is first seen at $t$ = 13.4 ms and persist throughout the post-FWI regime. The plateau occurs at cooler temperatures as time elapses. The temperature plateau suggests that gas temperatures have been resolved from the wall into the core-gas. When a distinct temperature plateau is observed, we argue that the limited temperature profile is sufficient to evaluate the thermal boundary layer.

In Fig. 4b, it is also evident that extrapolation of the gas temperature profile to the wall will not coincide with the measured wall temperature. This is particularly true during the burned-gas regime, which exhibit high gas temperatures. This aspect is not exclusive to this study, and has been shown in several studies utilizing CARS [23–25]. Beam steering is one possible explanation that describes this situation. Another possible explanation is that the flame/gas-wall interaction is highly transient; there is a limited amount of time that the wall is exposed to a flame/gas at a specific temperature. Due to this transient nature, there is insufficient time for the wall to respond to the hot gases compared to a steady-state situation where the flame/gas of a certain temperature remains fixed at the wall.

## 4.0 Results and discussion

### 4.1. Relevant boundary layer and heat transfer quantities

The thermal boundary layer development is analyzed in detail to understand the near-wall transient heat loss associated with polytropic compression and various burned-gas events. Scaling arguments are often used in boundary layer analyses, creating a self-similar solution with dimensionless units [12]. Such analysis requires velocity measurements, which are not available in this study. Therefore, in this study we analyze several quantities available from HRCARS and TGP to characterize the thermal boundary development and corresponding heat losses.

A key parameter to evaluate is the thermal boundary layer thickness, $\delta_T$. This can be defined by the thermal displacement thickness [12]:

$$\delta_T = \int_0^\infty \left( \frac{T_{core} - T(y)}{T_{core} - T_{wall}} \right) dy \qquad (1)$$

For all regimes other than the post-FWI (burned-gas) regime, $T_{core}$ is evaluated as the spatially-averaged temperature from $1.5 \leq y \leq 3$ mm, where the temperature profiles exhibit a uniform temperature distribution. As shown in Fig. 4b, for the post-FWI regime, temperature profiles do not span the entire 3 mm measurement length. Therefore, in the post-FWI regime, temperature profiles are fitted using Eqn. (2), where $A$, $B$, and $C$ are coefficients.

$$T(y) = A \exp(-By) + C \qquad (2)$$

All 105 temperature profiles in the post-FWI regime fit very well to Eqn. (2), with mean $R^2 \geq 0.97$ (1σ = 0.017). Using Eqn. (2), $T_{core}$ is set equal to the coefficient $C$. The upper limit for the value of $C$ is specified as the adiabatic flame temperature ($T_{ad}$ = 2,134 K). For temperature profiles which distinctly exhibit $T > 2,000$ K away from the wall (e.g. temp. profile at $t$ = 13.4 ms in Fig. 4b), the constraint $T_{core} = C = T_{ad}$ is applied. For temperature profiles exhibiting data end-points well below 2000 K (e.g. temp. profile at $t$ = 60.4 ms in Fig. 4b), $C$ is determined from Eqn. (2) (i.e., $T_{core} = C \neq T_{ad}$). This procedure provides a reasonable estimate of $T_{core}$ given the measurement limitations.

Several temperatures are also evaluated to characterize the boundary layer. $T_{core}$ and $T_{wall}$ are evaluated because these temperature reservoirs are responsible for the boundary layer. The spatially-averaged gas temperature from 50 μm to 150 μm ($\bar{T}_{gas,(y=50-150\mu m)} = T_{gas(nw)}$) is evaluated to provide a representative gas temperature within the boundary layer. The average thermal gradient ($\overline{\Delta T/\Delta y}$), calculated from 50µm to $\delta_T$, is also evaluated. $\overline{\Delta T/\Delta y}$ is calculated using a 5-pt moving average filter applied to the data.

The wall heat flux is calculated to quantify the wall heat loss.

$$\dot{Q}_{wall} = \left|-\lambda_{T_\lambda} \cdot \Delta T/\Delta y\right| \quad [kW/m^2] \qquad (3)$$

The thermal conductivity, $\lambda$, is determined at temperature $T_\lambda$, which is an average between the wall temperature and the spatially-averaged gas temperatures closest to the wall:

$$T_\lambda = \left(T_{wall} + \overline{T}_{gas(y=50,70,90\mu m)}\right)/2 \qquad (4)$$

The temperature gradient $\Delta T/\Delta y$ in Eqn. (3) is evaluated as:

$$\Delta T/\Delta y = \left(\overline{T}_{gas(y=50,70,90\mu m)} - T_{wall}\right)/\left(\overline{y}_{gas(y=50,70,90\mu m)} - y_{wall(y=0)}\right) \qquad (5)$$

Where, $\Delta y = \left(\overline{y}_{gas(y=50,70,90\mu m)} - y_{wall(y=0)}\right) = 70\mu m$.

To characterize the relative energy lost at the wall ($E_{loss,\%}$), the wall heat loss is compared to the change of internal energy of the core-gas:

$$E_{loss,\%} = (Q_{loss}/\Delta E_{core}) \times 100\% \qquad (6)$$

$$Q_{loss} = \dot{Q}_{wall} A_{s,u/b} \Delta t \quad [kJ] \qquad (7)$$

$$\Delta E_{core} = \rho V_{u/b} c_v (T_{core} - T_{initial}) \quad [kJ] \qquad (8)$$

$A_s$ represents the surface area of interest at time $t$. The surface area comprises of the top and bottom surfaces of the test-section. The *crevice* surface area is not considered since measurements were not performed in the crevice region, and temperature distributions will be different within the crevice. The subscript $u/b$ refers to unburned- or burned-gas depending on whether $E_{loss,\%}$ is evaluated during the unburned-gas or the burned-gas regime. For simplicity, $\Delta t$ is evaluated as 1 ms (0.001 sec), such that $Q_{loss}$ and $E_{loss,\%}$ evaluate the energy loss per ms. In Eqn. (8), $\rho$ is the gas density, $V$ is the unburned- or burned-gas volume in the test-section at time $t$, and $c_v$ is the specific heat at constant volume. $T_{initial}$ is the gas temperature before ignition (300 K).

CH* images are used to estimate $A_{s,u/b}$ and $V_{u/b}$. Imaging through the side window provides an estimate of the unburned- and burned-gas regions in the central *x-y* plane. The flame is assumed to propagate in a spherical fashion within the confines of the FVC. CH* images through the front window support this claim and are used to determine the extent of the enflamed region in the *z*-domain. While CH* images through the side window show a mild flame asymmetry in the *y*-direction, this slight deviation from the assumed spherical shape is minor within the calculation of $A_{s,u/b}$ and $V_{u/b}$.

The calculation of $E_{loss,\%}$ requires simplifying assumptions. As such, $E_{loss,\%}$ values are considered as estimations. A major underlying assumption is that HRCARS and TGP measurements are representative of all unburned (burned) gases within the test-section at time $t$. This assumption is required because measurements are only performed at a specific location. While spatial deviations can exist, gas temperature variations are expected to be largest in the *y*-direction, i.e. $\partial T/\partial y \gg \partial T/\partial x, \partial T/\partial z$. We also assume that the core-gas region has a homogeneous temperature distribution, and is captured within the 3 mm domain. 1D temperature profiles exhibit a uniform temperature region well outside $\delta_T$, supporting this assumption. It is not until late within the burned-gas regime when $T_{core}$ exhibits a heterogenous distribution (see Sect. 4.3.2), at which point $E_{loss,\%}$ is not calculated.

The uncertainty associated with $E_{loss,\%}$ is estimated by evaluating the measurable uncertainty of variables $\dot{Q}_{wall}$ and $(T_{core} - T_{initial})$ in Eqns (7) and (8) respectively. The uncertainty of $\dot{Q}_{wall}$ in the unburned regime is ±1.09 kW/m², while the uncertainty of $\dot{Q}_{wall}$ between 8.4-10.4 ms, when maximum heat flux is recorded, is ± 29.2 kW/m². It then follows that, during the unburned gas regime, the mean value of $E_{loss,\%}$ is 5.8 % with an uncertainty of ±1.6%. At the onset of the post-FWI regime (8.4 – 10.4 ms) where the peak $E_{loss,\%}$ and $\dot{Q}_{wall}$ are recorded, the mean value of $E_{loss,\%}$ is 32.8% with an uncertainty of ±2.7%. Both $\dot{Q}_{wall}$ and $E_{loss,\%}$ uncertainties steadily decrease to ±2.0 kW/m² and ±0.4 % at the end of the burned gas regime at $t$ = 100 ms.

Figure 5 shows the time-history of the aforementioned quantities throughout the main experimental duration. The various regimes are highlighted in Fig. 5, and all quantities are evaluated within the individual regimes in the following sections. Quantities are not reported during the initial flame propagation (-35 ≤ $t$ ≤ - 22 ms) because the weak compression at that stage does not yield a significant increase in unburned-gas temperature. For clarity, $T_{core}$ is plotted alongside relevant temperatures (Fig. 5a) as well as $\delta_T$ (Fig. 5b) because it helps describes the behavior of these quantities.

*4.2 Unburned-gas regime*

The unburned-gas regime is characterized by a mild polytropic compression process resulting from flame expansion. The compression precipitates an increase in unburned-gas temperature, which initiates thermal boundary layer development from wall heat loss. Figure 6 describes this process in detail. Selected 1D-temperature profiles are shown in Fig. 6a to provide a visual representation of the thermal boundary layer development. The remaining subplots in Fig. 6 represent those in Fig. 5, but solely focus on the unburned-gas regime. Data for $\overline{\Delta T / \Delta y}$ and $E_{loss,\%}$ are only shown for $t$ > -10 ms. Before that time, thermal gradients are small and so is the value $T_{core} - T_{initial}$ evaluated in Eqn. (8). The HRCARS uncertainty (2.7 K) is large in relation to these values. This yields large scatter and inaccurate values within these quantities before $t$ = -10 ms.

During this mild polytropic compression, $T_{core}$ increases up to 30 K its initial value, while $T_{wall}$ does not register an increase in temperature. As $T_{core}$ deviates from $T_{wall}$ as time progresses, the boundary layer becomes more pronounced within the temperature profiles in Fig. 6a. From $t$ = -15 ms, gas temperatures closest to the wall $(T_{gas(nw)})$ begin to deviate from $T_{core}$ and exhibit up to 15 K difference from $T_{core}$ and $T_{wall}$. As the boundary layer forms, $\delta_T$ remains small (≤ 200 µm) and values exhibit a moderate amount of scatter. This scatter is largest for data before $t$ = -10 ms, when the measurement uncertainty is significant in comparison to $T_{core} - T_{wall}$. Despite this scatter, $\delta_T$ shows a mild increasing trend up to 200 µm as the boundary layer becomes more pronounced.

During this mild boundary layer development, $\overline{\Delta T / \Delta y}$ increases up to 30 K/mm and the wall heat flux increases up to $\dot{Q}_{wall}$ = 6 kW/m². $\dot{Q}_{wall}$ values beyond $t$ = -10 ms represent ~ 6% energy loss per ms at the wall $(\dot{E}_{loss,\%})$. This value is greater than the 3% reported in [27], because those calculations included the crevice region for variables $A_s$ and $V$ (see Eqn. (7) and Eqn. (8)). This demonstrates that even for a mild compression event, discernible heat losses exist at the wall, which leads to appreciable thermal stratification through the boundary layer. Under more realistic engine conditions with higher pressures and temperatures, energy losses are expected to be higher, which lead to a greater thermal stratification that penetrates into the core-gases as shown in [5–7].

*4.3 Burned-gas regime*

The burned-gas regime occurs after the flame has progressed through the HRCARS volume. This regime occurs from 8 < $t$ < 100 ms and includes several transient events, which consequently modify the boundary layer development. For that reason, the burned-gas regime is categorized into the following sub-regimes: (a) post-FWI, (b) outgassing, and (c) end-exhaust.

*4.3.1 post-FWI*

The post-FWI regime occurs directly after FWI and exists from $8 \leq t \leq 63$ ms. This regime includes the remaining compression and the first half of expansion. Figure 7 shows the data for the post-FWI regime. As mentioned in Sect. 3, 1D-temperature profiles do not span the entire 3 mm measurement length. It is not until $t = 13.4$ mm until data is resolved beyond $y = 1$ mm. However, as time progresses, more of the measurement length is resolved and the entire length is resolved near the end of the post-FWI regime. The lines shown for the selected temperature profiles shown in Fig. 7a represent the data fit according to Eqn. (2). The timing of peak pressure ($P_{peak}$) is also shown within the plots of Fig. 7.

During the first-half of the post-FWI regime, compression continues until $t = 36$ ms, when the peak pressure is reached. During this time, gas temperatures are the highest of those recorded as the flame just finished passing through the HRCARS volume. HRCARS data points furthest from the wall exhibit gas temperatures above 2,000 K. As described in Sect. 4.1, these temperatures are approximated as $T_{core} \sim T_{ad}$. This occurs for majority of data from $13 \leq t \leq 30$ ms, after which $T_{core}$ decreases once $P_{peak}$ is reached. $T_{wall}$ reaches its maximum value of 312 K direct after FWI at $t = 10$ ms. The maximum $T_{wall}$ is temporarily sustained and then $T_{wall}$ decreases rather quickly to 303 K near $t = 36$ ms. A slight oscillatory behavior is shown for $T_{wall}$. These oscillations occur as the post-flame products attenuate the phosphorescence signals, yielding lower SNR levels.

As $T_{wall}$ rapidly cools while $T_{core} \sim T_{ad}$, $\delta_T$ exhibits a substantial linear growth from 290 μm to 700 μm. This growth is qualitatively seen in the profiles in Fig. 7a. As $P_{peak}$ is reached, $\delta_T$ begins to level off near 730 μm. During this time ($13 \leq t \leq 30$ ms), $T_{gas(nw)}$ reaches up to 1250 K, and remains ~ 1200 K lower than $T_{ad}$. Average thermal gradients in the boundary layer are in excess of 2000 K/mm for $t <$ 15ms, but decrease rapidly to 1100 K/mm as $P_{peak}$ is reached. Similarly, $\dot{Q}_{wall}$ peaks at 650 kW/m² directly after FWI at $t = 10$ ms, and decreases rapidly to 260 kW/m² as $P_{peak}$ is reached. These $\dot{Q}_{wall}$ values are comparable with values reported for motored engine operation (100-800 kW/m²), but are below values reported under fired operation (800-5000 kW/m²) [2,20,38,39]. Peak energy losses ($E_{loss,\%}$) in excess of 30% occur as $\dot{Q}_{wall,max}$ is reached and $E_{loss,\%}$ decays in a similar fashion down to ~10% as $P_{peak}$ is reached.

The remaining portion of the post-FWI regime exhibits the expansion phase ($36 < t \leq 63$ ms) for which pressure decreases from $P = 2.08 – 1.53$ bar. During this time, $T_{core}$ decreases linearly from 1900 K to 1200 K, while $T_{wall}$ decreases more modestly from ~303 K to ~300 K. As $T_{core}$ cools more rapidly than $T_{wall}$, $\overline{\Delta T/\Delta y}$ begin to stabilize between 700-800 K/mm. These attributes are qualitatively shown in the selected temperature profiles in Fig. 7a. Consequently, $\delta_T$ shows a minimal increase and stabilizes near an average value of 740 μm, but exhibits a considerable amount of scatter. The scatter in $\delta_T$ arises from the larger temperature scatter beyond $y = 2$ mm due to lower SNR in those regions (see Sect. 3.1). From $36 < t \leq 63$ ms, $\dot{Q}_{wall}$ decays steadily from 250 – 100 kW/m², while $E_{loss,\%}$ decreases from 11% to 7%.

*4.3.2 Outgassing regime*

During expansion, gases leave the FVC through the dump-valve and gases from the crevice region slowly enter the test-section region. This event is termed "outgassing" [4–6]. During outgassing, significant thermal mixing takes place, which modifies the boundary layer development. Figure 8 describes the near-wall processes during the outgassing regime. Unlike previous figures, the quantities $\delta_T$ and $E_{loss,\%}$ and are not reported in Fig. 8. This is because of extreme changes in $T_{core}$ during outgassing, which yields unrealistic values of these quantities. The $T_{core}$ values reported in Fig. 8 are calculated as the spatially averaged temperature from $1.5 \leq y \leq 3$ mm. $\overline{\Delta T/\Delta y}$ values reported in Fig. 8 are calculated from $50 \leq y \leq 400$ μm.

The selected 1D-temperature profiles in Fig. 8a exhibit compelling differences in the boundary layer behavior. At $t$ = 63.2 ms, the temperature profile appears similar to those shown at the end of the post-FWI regime, where sharp temperature gradients extend from the wall and temperatures converge to a representative $T_{core}$ value from $1.5 \leq y \leq 2$ mm. However, at $t$ = 63.2 ms, a noticeable temperature decrease of ~200 K exists for gases beyond 2 mm. At $t$ = 68.2 ms, gas temperatures away from the wall decrease more significantly as an indisputable temperature reversal exists with gas temperatures decreasing from 1200 K to 400 K. These temperature reversals are prevalent throughout the outgassing regime and completely disrupt the pre-existing boundary layer. As time elapses from $t$ = 68.2 ms, the temperature reversals occur closer to the wall and profiles suggest that considerable thermal mixing occurs within the core-gas region. Near the end of the outgassing regime, 1D-temperature profiles show a resemblance of the pre-existing thermal boundary layer, albeit at lower temperatures with mild temperature variations in the core-gas.

Figure 8b shows variations in $T_{core}$ due to these temperature reversals. These variations are most severe from $t$ = 63 – 70 ms after which $T_{core}$ stabilizes between 450-550 K for the remainder of this regime. $\overline{\Delta T/\Delta y}$ shows similar gradient values from $t$ = 63 – 73, after which the temperature reversals near the wall reduce $\overline{\Delta T/\Delta y}$ substantially. $\dot{Q}_{wall}$ along with $T_{gas(nw)}$ exhibit a mild increase from $t$ = 63 – 71 ms, after which both quantities show a clear decreasing trend. The mild increases in $\dot{Q}_{wall}$ and $T_{gas(nw)}$ may be a result of chemical reactions from unburned gas near the wall (see below), but require further measurements to confirm. $T_{wall}$ values are within 297 – 300 K and exhibit the same oscillatory behavior as in Fig. 7b.

The gas temperature behavior in the outgassing regime is a consequence of unburned gases released from the crevice region during expansion. To substantiate this hypothesis, we take advantage of $O_2/N_2$ measurements from HRCARS. Our HRCARS model requires further $O_2$ linewidth information to provide $O_2/N_2$ concentrations at the same uncertainty levels as gas temperature. While uncertainties are higher for $O_2/N_2$, we utilize such measurements to present compelling findings during outgassing. Figure 9 shows selected temperature and $O_2/N_2$ profiles shortly before and during outgassing. At $t$ = 59.3 ms, before outgassing, a 'traditional' boundary layer profile is shown extending to 1250 K away from the wall. The $O_2/N_2$ ratio is low (< 0.05) for the entire profile. During outgassing, at $t$ = 73.2 ms, the temperature profile below $y$ = 0.5 mm is similar to that shown at 59.6 ms, and the $O_2/N_2$ ratio remains below 0.08. However, above $y$ = 0.5 mm a strong temperature reversal exists and gas temperatures as low as 400 K exist at $y$ = 1.3 mm. To put this into perspective, the gas temperature at $y$ = 1.3 mm is ~ 50 K colder than the gas temperature at the wall. As the temperature decreases from $0.5 \leq y \leq 1.3$ mm, $O_2/N_2$ ratios increase to 0.2, which resembles ratios expected within the unburned-gas. Above $y$ = 1.3 mm, the temperature gradually increases up to 600 K, while $O_2/N_2$ decreases to 0.14.

Indeed, the additional $O_2/N_2$ information help describe the physical processes responsible for the observed temperature behavior. The profile at $t$ = 72.3 ms in Fig. 9, suggest that a local pocket, seemingly comprised of mostly unburned gas (400 K, 0.2 $O_2/N_2$), passes through the HRCARS volume near $y$ = 1.3 mm. The increase in temperature and decrease in $O_2/N_2$ above $y$ = 1.3 mm suggests a degree of thermal mixing between unburned and burned gases in the core-gas region.

The CH* images visualizing the enflamed regions in the crevice demonstrate that a large volume of unburned gases can exist in the crevice (see Fig. 2). These gases exit the crevice during expansion. Under the current operating conditions ($CH_4$/air, 1.02 bar initial pressure), chemiluminescence images within the test-section do not provide additional insight of the outgassing event, as signals are too weak during this time. However, gases exiting the crevice are more visibly seen under alternative operating conditions. Figure 10 shows burned-gas luminescence images for experiments utilizing $C_2H_4$ and 2.03 bar initial pressure. These operating conditions yield higher signal intensity and a larger pressure drop from $P_{peak}$, which are more conducive to visualize gas release from the crevice. The reference time in Fig. 10 is referenced to valve opening (Vo). The image sequence shows the progression of a lower intensity region discharging from the crevice region and protruding into the test-section. Unburned- and flame-quenched gases will have lower intensity as they will have little-to-no luminescence signal. These images further support the hypothesis that colder gas temperatures during the outgassing regime represent gases emanating from the crevice region.

*4.3.3 End-exhaust regime*

At the end of the outgassing regime, temperature profiles transition back into the profile shapes seen before outgassing. Since these profiles exhibit different characteristics than during outgassing, we classify a new regime in which the boundary layer is analyzed. This regime is called the end-exhaust regime, which is comprised of latter end of expansion/exhaust from $83 < t \leq 100$ ms.

Figure 11 describes the boundary layer development and wall heat loss within this regime. 1D-temperature profiles in Fig. 11a resemble those similar to post-FWI, except that temperatures are now much lower and profiles continue to exhibit temperature reversals away from the wall. These temperature reversals are small compared to outgassing and are more prevalent for earlier reference times indicating that thermal mixing persists in the core-gas. Gas temperatures decrease and the temperature reversals subside as time elapses. $T_{core}$ values in Fig. 11b are represented as the spatially-averaged temperature from $1.5 \leq t \leq 3$ mm. $T_{core}$ exhibits a gradual decay from 550 K to 500 K. These values are lower than the highest temperatures within the 1D profiles due to the temperature reversals. $T_{wall}$ exhibits a quasi-steady value near 297 K. $T_{gas(nw)}$ remains ~ 100 K lower than $T_{core}$ and exhibit a similar decreasing trend.

Due to the evolving temperature reversals, $\delta_T$ will contain inconsistent values compared to previous regimes. $\delta_T$ exhibit considerable scatter and are underestimated for profiles exhibiting temperature reversals. Due to the inconsistent calculation of $\delta_T$, $\overline{\Delta T/\Delta y}$ values are calculated from $50 \leq y \leq 400$ μm within the end-exhaust regime. $\overline{\Delta T/\Delta y}$ show a mild decrease within the data scatter. The wall heat flux, $\dot{Q}_{wall}$, also exhibits a mild decrease from 50 to 25 kW/m². The corresponding $\dot{Q}_{wall}$ and $T_{core}$ values register energy losses from 5-8% at the wall. Although $E_{loss,\%}$ values are lower than during post-FWI, they demonstrate the continued energy loss well within expansion/exhaust, which for real engines, represent a lost opportunity for work.

Figure 11 indicates that thermal processes continue beyond $t$ = 100 ms as temperatures have not reached equilibrium. Beyond $t$ = 100 ms, temperatures continue to decay, and shortly afterwards the chamber is evacuated to conclude the experiment.

*4.4 Comparison to canonical environments*

Although the FVC is an overly simplified environment of a piston engine, it is also fundamentally different from canonical environments where boundary layers have been studied more extensively. For example, the outer gases in the FVC are not constant in pressure, temperature or velocity. The boundary layer is unsteady and contains chemical reactions, which both deviate from the traditional, steady-state conditions within canonical environments. In this section, the thermal boundary layer development in the FVC is compared to those studied in a flow over a flat plate [40]. This comparison is useful in order to characterize the FVC in relation to well-known conditions. This canonical configuration is chosen because it is often the foundation of wall-models for engines [17,18].

In this comparison, we analyze normalized temperature profiles, defined as:

$$\theta(y) = (T(y) - T_{wall})/(T_{core} - T_{wall}) \tag{9}$$

Figure 12a shows individual normalized temperature profiles for selected reference times during relevant regimes. The normalized temperature profile for a steady-state *laminar* and *turbulent* forced flow over a flat plate from Siebers et al.[40] are also shown. As observed in previous studies [23,40], profiles within the highlighted region from $-0.4 \leq log_{10}(y/\delta_T) \leq 0.5$ exhibit a quasi-linear behavior. This region often pertains to the buffer and outer regions of turbulent boundary layers [23,41]. In our work, the lack of velocity data prevent us from identifying the buffer layer region. During compression, the flame itself is laminar, but conditions arguably change during expansion. In Fig. 12b, the $-0.4 \leq log_{10}(y/\delta_T) \leq 0.5$ region is highlighted for the individual temperature profiles and corresponds to a region of $0.4 \leq y/\delta_T \leq 3.2$. The slope from $-0.4 \leq log_{10}(y/\delta_T) \leq 0.5$ is calculated:

$$z_{\delta/\theta} = \Delta(log_{10}(y/\delta_T))/\Delta\theta(y) \tag{10}$$

$z_{\delta/\theta}$ is shown in Fig. 12c for all measurement points to describe its evolution throughout the various regimes. Values for $z_{\delta/\theta}$ are also shown for the laminar and turbulent forced flow data from [40]. Additionally, $z_{\delta/\theta}$ values are shown for an *unsteady* laminar flow over a flat plate from Ojo et al. [42]. The data from [40,42] only consider non-reacting flows. Smaller values of $z_{\delta/\theta}$ indicate larger changes in $\theta(y)$ (i.e. boundary layer profile) from $0.4 \leq y/\delta_T \leq 3.2$. Larger $z_{\delta/\theta}$ values indicate smaller changes in $\theta(y)$. This analysis is intended to describe the boundary layer shape relative to laminar and turbulent flows over a flat plate.

During the unburned-gas regime, $z_{\delta/\theta}$ values progressively increase to values in between the laminar and turbulent values from Siebers et al [40]. Values below 0.6 correspond to data with small temperature changes (< 10 K), when a boundary layer is first initiated. As the boundary layer becomes more pronounced during compression, the temperature profile changes less significantly, yielding higher $z_{\delta/\theta}$ values. During the post-FWI regime, $z_{\delta/\theta}$ remains relatively constant and in close agreement with the unsteady, laminar data from Ojo et al. [42]. During post-FWI, temperature profiles exhibit larger changes from $0.4 \leq y/\delta_T \leq 3.2$ (e.g. see Fig. 12b), which result in the lower $z_{\delta/\theta}$ values. Due to strong temperature reversals, $\delta_T$ (and thus $z_{\delta/\theta}$) is not calculated during the outgassing regime from $t$ = 63-83 ms. During the end-exhaust regime, $z_{\delta/\theta}$ increases as changes in temperature profiles from $0.4 \leq y/\delta_T \leq 3.2$ become less pronounced. The scatter during the end-exhaust regime arises from temperature reversals still present after outgassing. These less homogeneous temperature distributions yield larger fluctuations in $\delta_T$.

This analysis demonstrates the transient nature of the thermal boundary layer for engine-relevant processes. These analyses presented within the results are intended to provide useful measurements to guide numerical modelling for developing improved heat transfer models for engines.

### 5.0    Conclusion

This work investigates the thermal boundary layer development and corresponding wall heat loss associated with engine-relevant processes. We take advantage of HRCARS to measure single-shot, spatially-resolved gas temperatures, which provide exclusive access to the thermal boundary layer. Phosphor thermometry is used to measure wall temperature and assess wall heat fluxes. Measurements are conducted in a FVC that emulates an engine geometry and simulates mild compression and expansion processes at engine timescales. This simplified environment is conducive for fundamental boundary layer and heat transfer studies associated with engine-relevant processes.

Assessment of HRCARS to resolve the thermal boundary layer is performed. Gas temperature are confined within a 3 mm length normal to a surface. Temperatures away from the wall reach a converged value, indicating that core-gases outside the boundary layer are resolved within the 3 mm length. SNR levels are lowest from $y$ = 2-3 mm, causing larger scatter in that region. After FWI, beam steering and low SNR prevent temperature measurements through the entire 3 mm length. From $t$ = 13.4 ms, measurements extend beyond $y$ = 1 mm and temperatures furthest from the wall exhibit converged values near $T_{ad}$. While temperatures are not resolved in the entire 3 mm, measurements are suitable to evaluate the thermal boundary layer.

The evolution of integral boundary layer and heat loss quantities are evaluated during the unburned-gas and burned-gas regime. The burned-gas regime is categorized into sub-regimes to evaluate quantities associated with distinct events during expansion. During the unburned-gas regime, the mild compression process yields an increase of $T_{core}$ by 30 K, while $T_{wall}$ remains unchanged. A thermal boundary layer is initiated during compression and $\delta_T$ increases up to 200 μm. Wall heat fluxes remain small ($\leq$ 6 kW/m$^2$) during this regime, but correspond to ~6% relative energy loss per ms.

During the post-FWI regime, $T_{core}$ values are the highest recorded, while $T_{wall}$ increases by 16 K. The large difference between $T_{core}$ and $T_{wall}$ leads to a rapid increase in $\delta_T$ from 290-730 μm. Average thermal gradients in excess of 2000 K/mm exist during this time. The maximum wall heat flux (650 kW/m$^2$) is reached and yields $E_{loss,\%}$ values in excess of 25%. As time progresses and expansion begins, $T_{core}$ decreases more substantially than $T_{wall}$ for which $\delta_T$ stabilizes between 600 – 800 μm. Values of $\dot{Q}_{wall}$ and $E_{loss,\%}$ decrease monotonically during expansion, but demonstrate that relative energy losses between 10-20% persist during expansion.

During mid-expansion, distinct temperature reversals are observed in the 1D-temperature profiles, which defines the outgassing regime. O$_2$/N$_2$ measurements from HRCARS demonstrate that these temperature reversals occur as colder fresh-gases from the crevice pass through the HCARS volume and thermally mix with the hotter burned-gases away from the wall. HRCARS appropriately resolve these transient thermal mixing processes, which penetrate into the boundary layer.

Normalized temperature profiles are analyzed and compared to canonical boundary layers. This analysis is intended to describe the thermal boundary layer shape relative to laminar and turbulent conditions for a flow over a flat plate. Findings reveal that the boundary layer shape spans regions between laminar and turbulent flows over a flat plate.

This work demonstrates the transient nature of the thermal boundary layer and associated energy losses at the wall during these engine-relevant processes. These processes are important to predict within engine simulations in order to serve as accurate tools to study relevant in-cylinder processes underpinning engine performance.

**Acknowledgements**

We gratefully acknowledge funding from the European Research Council (grant #759549) and EPSRC (EP/P020593/1).**References**

[1]   A. Dreizler, B. Böhm, Advanced laser diagnostics for an improved understanding of premixed flame-wall interactions, Proc. Combust. Inst. 35 (2015) 37–64. https://doi.org/https://doi.org/10.1016/j.proci.2014.08.014.

[2]   G. Borman, K. Nishiwaki, Internal-combustion engine heat transfer, Prog. Energy Combust. Sci. 13 (1987) 1-46. https://doi.org/10.1016/0360-1285(87)90005-0.

[3]   J.P. Szybist, S. Busch, R.L. McCormick, J.A. Pihl, D.A. Splitter, M.A. Ratcliff, et al, What fuel properties enable higher thermal efficiency in spark-ignited engines?, Prog. Energy Combust. Sci. 82 (2021) 100876. https://doi.org/10.1016/j.pecs.2020.100876.

[4]   B. Peterson, E. Baum, B. Böhm, V. Sick, A. Dreizler, High-speed PIV and LIF imaging of temperature stratification in an internal combustion engine, Proc. Combust. Inst. 34 (2013) 3653–60. https://doi.org/http://dx.doi.org/10.1016/j.proci.2012.05.051.

[5]   N. Dronniou, J.E. Dec, Investigating the Development of Thermal Stratification from the Near-Wall Regions to the Bulk-Gas in an HCCI Engine with Planar Imaging Thermometry,

SAE Int J Engines 5 (2012) 1046–74. https://doi.org/10.4271/2012-01-1111.

[6] B. Peterson, E. Baum, B. Böhm, V. Sick, A. Dreizler, Evaluation of toluene LIF thermometry detection strategies applied in an internal combustion engine, Appl Phys B 117 (2014) 151–75. https://doi.org/10.1007/s00340-014-5815-0.

[7] S.A. Kaiser, M. Schild, C. Schulz, Thermal stratification in an internal combustion engine due to wall heat transfer measured by laser-induced fluorescence, Proc. Combust. Inst. 34 (2013) 2911–9. https://doi.org/10.1016/j.proci.2012.05.059.

[8] M.K. Alzuabi, A. Wu, V.Sick, Experimental and numerical investigation of temperature fluctuations in the near-wall region of an optical reciprocating engine, Proc. Combust. Inst. 38 (2021) 5879-87. https://doi.org/10.1016/j.proci.2020.08.062.

[9] J.E. Dec, W. Hwang, Characterizing the Development of Thermal Stratification in an HCCI Engine Using Planar-Imaging Thermometry, SAE Int J Engines 2 (2009) 421–38. https://doi.org/10.4271/2009-01-0650.

[10] B. Peterson, E. Baum, B. Böhm, V. Sick, A. Dreizler, Spray-induced temperature stratification dynamics in a gasoline direct-injection engine, Proc. Combust. Inst. 35 (2015) 2923–31. https://doi.org/10.1016/j.proci.2014.06.103.

[11] M. Schmitt, C.E. Frouzakis, Y.M. Wright, A.G. Tomboulides, K.Boulouchos, Direct numerical simulation of the compression stroke under engine-relevant conditions: Evolution of the velocity and thermal boundary layers, Int. J. Heat Mass Transf. 91 (2015) 948–60. https://doi.org/10.1016/j.ijheatmasstransfer.2015.08.031.

[12] H. Schlichting, K. Gersten, Boundary-layer Theory, Springer Berlin, 2017.

[13] A.Y. Alharbi, V Sick, Investigation of boundary layers in internal combustion engines using a hybrid algorithm of high speed micro-PIV and PTV, Exp. Fluids 49 (2010) 949–59. https://doi.org/10.1007/s00348-010-0870-8.

[14] C. Jainski, L. Lu, A. Dreizler, V.Sick, High-speed micro particle image velocimetry studies of boundary-layer flows in a direct-injection engine, Int. J. Engine Res. 14 (2012) 247–59. https://doi.org/10.1177/1468087412455746.

[15] C.P. Ding, B. Peterson, M. Schmidt, A. Dreizler, B. Böhm, Flame/flow dynamics at the piston surface of an IC engine measured by high-speed PLIF and PTV, Proc. Combust. Inst. 37 (2019) 4973–81. https://doi.org/https://doi.org/10.1016/j.proci.2018.06.215.

[16] M Schmidt, C.P. Ding, B. Peterson, A. Dreizler, B. Böhm, Near-Wall Flame and Flow Measurements in an Optically Accessible SI Engine, Flow Turbul. Combust., 106 (2021), 597-611. https://doi.org/10.1007/s10494-020-00147-9.

[17] A. Renaud, C.P. Ding, S. Jakirlic, A. Dreizler, B. Böhm, Experimental characterization of the velocity boundary layer in a motored IC engine, Int. J. Heat Fluid Flow 71 (2018) 366–77. https://doi.org/10.1016/j.ijheatfluidflow.2018.04.014.

[18] P.C. Ma, T. Ewan, C. Jainski, L. Lu, A. Dreizler, V, Sick, et al., Development and Analysis of Wall Models for Internal Combustion Engine Simulations Using High-speed Micro-PIV Measurements, Flow Turbul. Combust., 98 (2017) 283–309. https://doi.org/10.1007/s10494-016-9734-5.

[19] M. Schmitt, C.E. Frouzakis, Y.M. Wright, A.G. Tomboulides, K.Boulouchos, Investigation of wall heat transfer and thermal stratification under engine-relevant conditions using DNS, Int. J. Engine Res., 17 (2015) 63–75. https://doi.org/10.1177/1468087415588710.

[20] P.C. Ma, M. Greene, V. Sick, M. Ihme, Non-equilibrium wall-modeling for internal combustion engine simulations with wall heat transfer, Int. J. Engine Res., 18 (2017) 15–25.

https://doi.org/10.1177/1468087416686699.

[21] S.P. Kearney, R.P. Lucht, A.M. Jacobi, Temperature measurements in convective heat transfer flows using dual-broadband, pure-rotational coherent anti-Stokes Raman spectroscopy (CARS), Exp. Therm. Fluid Sci., 19 (1999) 13–26. https://doi.org/https://doi.org/10.1016/S0894-1777(99)00004-7.

[22] H. Kosaka, F. Zentgraf, A. Scholtissek, L. Bischoff, T. Häber, R. Suntz, et al., Wall heat fluxes and CO formation/oxidation during laminar and turbulent side-wall quenching of methane and DME flames, Int. J. Heat Fluid Flow, 70 (2018) 181–92. https://doi.org/https://doi.org/10.1016/j.ijheatfluidflow.2018.01.009.

[23] R.P. Lucht, D. Dunn-Rankin, T. Walter, T. Dreier, S.C. Bopp, Heat Transfer in Engines: Comparison of Cars Thermal Boundary Layer Measurements and Heat Flux Measurements†., Int. Congr. Expo., SAE International, SAE Technical Paper 910722 (1991). https://doi.org/https://doi.org/10.4271/910722.

[24] B. Grandin, I. Denbratt, J. Bood, C. Brackmann, P.E. Bengtsson, The Effect of Knock on the Heat Transfer in an SI Engine: Thermal Boundary Layer Investigation using CARS Temperature Measurements and Heat Flux Measurements, Int. Fuels Lubr. Meet. Expo., SAE International, SAE Technical Paper 2000-01-2831, (2000). https://doi.org/https://doi.org/10.4271/2000-01-2831.

[25] A. Bohlin, M. Mann, B.D. Patterson, A. Dreizler, C.J. Kliewer, Development of two-beam femtosecond/picosecond one-dimensional rotational coherent anti-Stokes Raman spectroscopy: Time-resolved probing of flame wall interactions, Proc. Combust. Inst., 35 (2015) 3723-30. https://doi.org/10.1016/j.proci.2014.05.124.

[26] A. Bohlin, C. Jainski, B.D. Patterson, A. Dreizler, C.J. Kliewer, Multiparameter spatio-thermochemical probing of flame–wall interactions advanced with coherent Raman imaging, Proc. Combust. Inst. 36 (2017) 4557–64. https://doi.org/10.1016/j.proci.2016.07.062.

[27] D. Escofet-Martin, A.O. Ojo, N.T. Mecker, M.A. Linne, B. Peterson, Simultaneous 1D hybrid fs/ps rotational CARS, phosphor thermometry, and CH* imaging to study transient near-wall heat transfer processes, Proc. Combust. Inst. 38 (2021) 1579-87. https://doi.org/https://doi.org/10.1016/j.proci.2020.06.097.

[28] A. Bohlin, B.D. Patterson, C.J. Kliewer, Communication: Simplified two-beam rotational CARS signal generation demonstrated in 1D, J Chem. Phys. 138 (2013) 81102. https://doi.org/10.1063/1.4793556.

[29] W.D. Kulatilaka, H.U. Stauffer, J.R. Gord, S. Roy, One-dimensional single-shot thermometry in flames using femtosecond-CARS line imaging, Opt. Lett. 36 (2011) 4182–4. https://doi.org/10.1364/OL.36.004182.

[30] J.E. Retter, G.S. Elliott, S.P. Kearney, Dielectric-barrier-discharge plasma-assisted hydrogen diffusion flame. Part 1: Temperature, oxygen, and fuel measurements by one-dimensional fs/ps rotational CARS imaging, Combust. Flame, 191 (2018) 527-540. https://doi.org/10.1016/j.combustflame.2018.01.031.

[31] C.J. Kliewer, Y. Gao, T. Seeger, B.D. Patterson, R.L. Farrow, T.B. Settersten, Quantitative one-dimensional imaging using picosecond dual-broadband pure-rotational coherent anti-Stokes Raman spectroscopy, Appl. Opt. 50 (2011) 1770–8. https://doi.org/10.1364/AO.50.001770..

[32] D. Escofet-Martin, A.O. Ojo, J. Collins, N.T. Mecker, M. Linne, B. Peterson, Dual-probe 1D hybrid fs/ps rotational CARS for simultaneous single-shot temperature, pressure, and $O_2/N_2$ measurements, Opt. Lett. 45 (2020) 4758–61. https://doi.org/10.1364/OL.400595.

[33]  M.A. Linne, R.K. Mackay, W.L. Bahn, A blow-down combustion bomb for engine in-cylinder kinetics research, Rev. Sci. Instrum. 65 (1994) 3834. https://doi.org/10.1063/1.1145173.

[34]  C.J. Kliewer, A. Bohlin, E. Nordström, B.D. Patterson, P.E. Bengtsson, T.B. Settersten, Time-domain measurements of S-branch $N_2$-$N_2$ Raman linewidths using picosecond pure rotational coherent anti-Stokes Raman spectroscopy, Appl. Phys. B Lasers Opt. 108 (2012) 419–426. https://doi.org/10.1007/s00340-012-5037-2.

[35]  G. Millot, R. Saint-Loup, J. Santos, R. Chaux, H. Berger, J, Bonamy, Collisional effects in the stimulated Raman Q branch of $O_2$ and $O_2$-$N_2$, J Chem. Phys. 96 (1992) 961-971. https://doi.org/10.1063/1.462116.

[36]  N. Fuhrmann, C. Litterscheid, C.P. Ding, J. Brübach, B. Albert, A. Dreizler, Cylinder head temperature determination using high-speed phosphor thermometry in a fired internal combustion engine, Appl. Phys. B Lasers Opt. 2 (2014) 293-303. https://doi.org/10.1007/s00340-013-5690-0.

[37]  P. Weigand, R. Lückerath, W.Meier, Documentation of flat premixed laminar $CH_4$/air standard flames: temperatures and species concentrations, http//Www Dlr de/VT/Datenarchiv (2003).

[38]  C.D. Rakopoulos, G.M. Kosmadakis, E.G. Pariotis, Critical evaluation of current heat transfer models used in CFD in-cylinder engine simulations and establishment of a comprehensive wall-function formulation, Appl. Energy 87 (2010) 1612–30. https://doi.org/https://doi.org/10.1016/j.apenergy.2009.09.029.

[39]  Z, Han, R.D. Reitz, A temperature wall function formulation for variable-density turbulent flows with application to engine convective heat transfer modeling, Int. J Heat Mass Transf. 40 (1997) 613–25. https://doi.org/10.1016/0017-9310(96)00117-2.

[40]  D.L. Siebers, R.G. Schwind, R.J. Moffatt, Experimental mixed convection heat transfer from a large, vertical surface in a horizontal flow, No. SAND83-8225, Sandia National Lab. (SNL-CA), Livermore, CA. United States: (1983).

[41]  R.B. Bird, W.E. Stewart, E.N. Lightfoot, Transport Phenomena, John Wiley, New York, 1960.

[42]  A.O. Ojo, B. Fond, C. Abram, B.G. Van Wachem, A.L. Heyes, F. Beyrau, Simultaneous measurements of the thermal and velocity boundary layer over a heated flat plate using thermographic Laser Doppler Velocimetry, International Heat Transfer Conference Digital Library, Begel House Inc. (2018) 8775–80. https://doi.org/10.1615/IHTC16.tpm.023986.

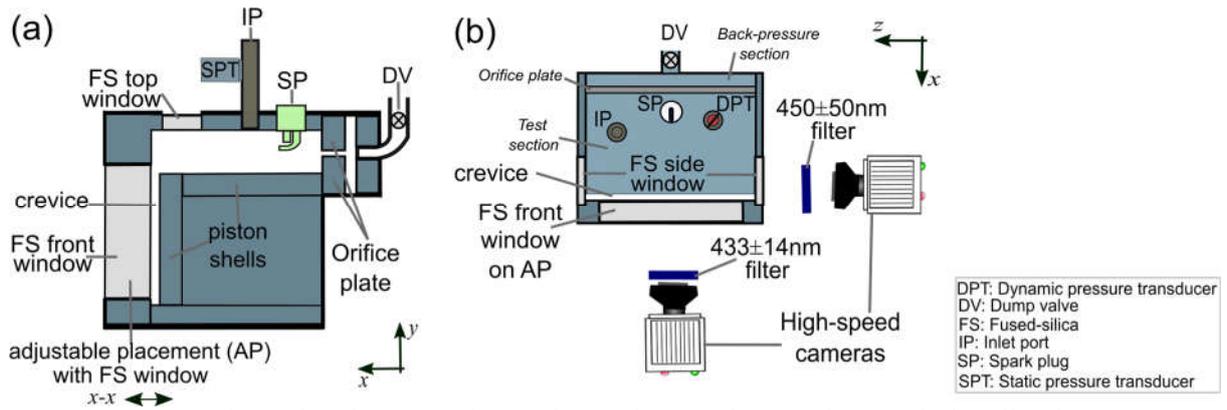

**Figure 1:** Schematic of FVC and experimental setup for supplemental chemiluminescence images to describe flame propagation in the test-section and crevice region. (a) side view, (b) top view.

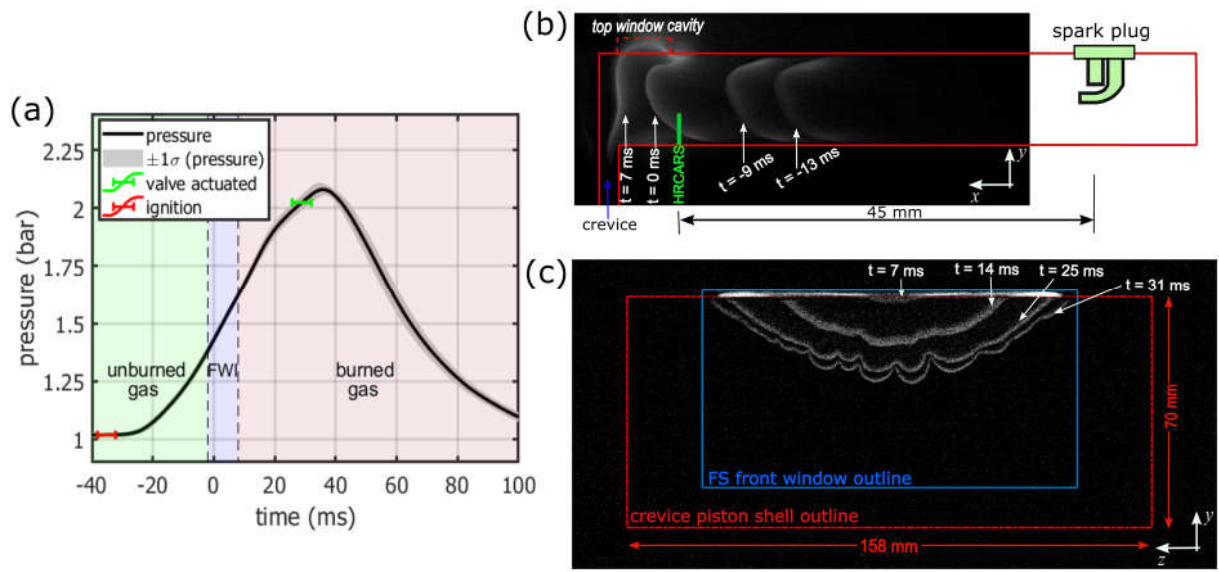

**Figure 2:** (a) pressure curve for methane/air (Φ = 0.9) and initial pressure of 1.02 bar. (b) and (c) CH* images showing flame propagation at selected times within the test-section and crevice region, respectively. Red dash-dot outline represents the piston shell (Δz × Δy : 158×70 mm) of the crevice region. Blue outline represents boundary of the Fuse-silica front window (Δz × Δy : ~ 110×60 mm). Crevice spacing is 2 mm for (b) and (c) to improve CH* signals in the crevice region. Trends are similar to 1.25 mm crevice spacing.

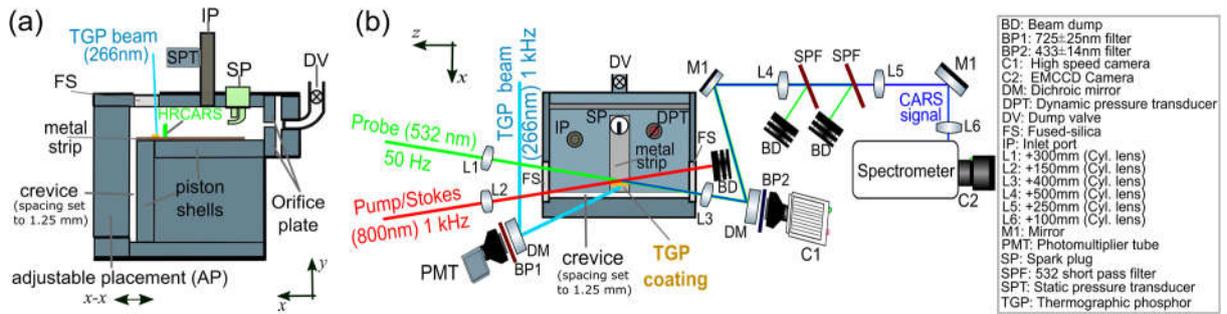

**Figure 3:** Schematic of experimental setup for HRCARS, TGP, and CH* imaging (a) side view of FVC showing HRCARS and TGP measurement locations, (b) top view of FVC.

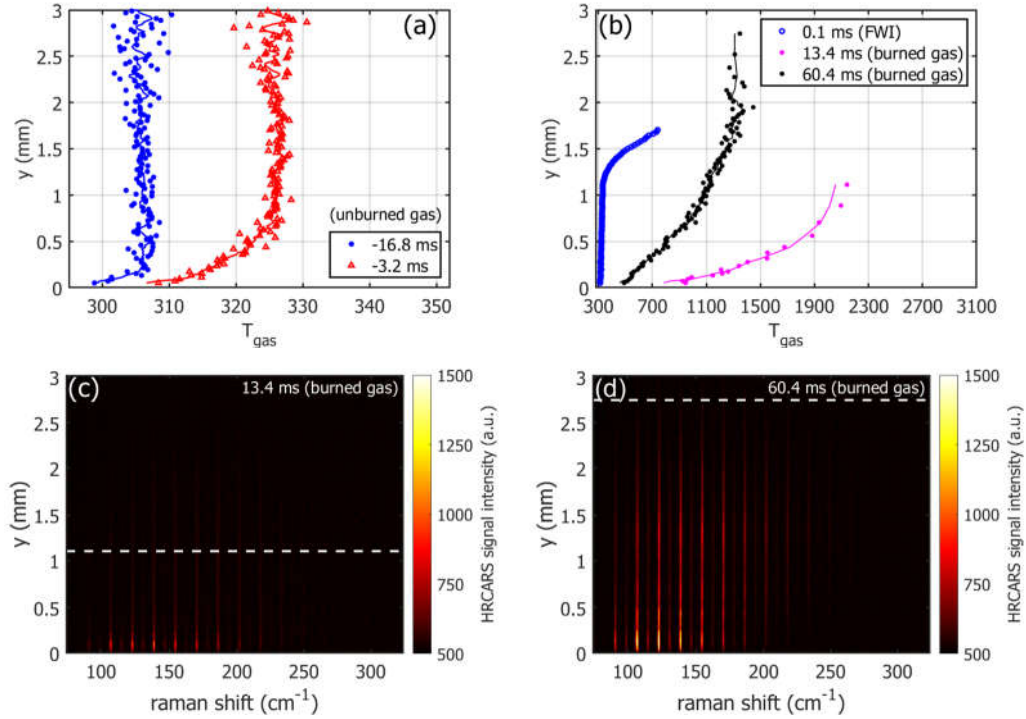

**Figure 4:** Selected 1D gas temperature profiles in (a) unburned-gas regime, (b) FWI and burned-gas (post-FWI) regimes. Lines plotted through the gas temperature profiles in (a) and (b) are 5-point moving average (100 μm) to guide the eye. (c) single-shot 1D HRCARS spectra at 13.4 ms, (d) single-shot 1D HRCARS spectra at 60.4 ms. Locations above white dashed lines in (c) and (d) contain SNR < 15, for which temperatures are not reported.

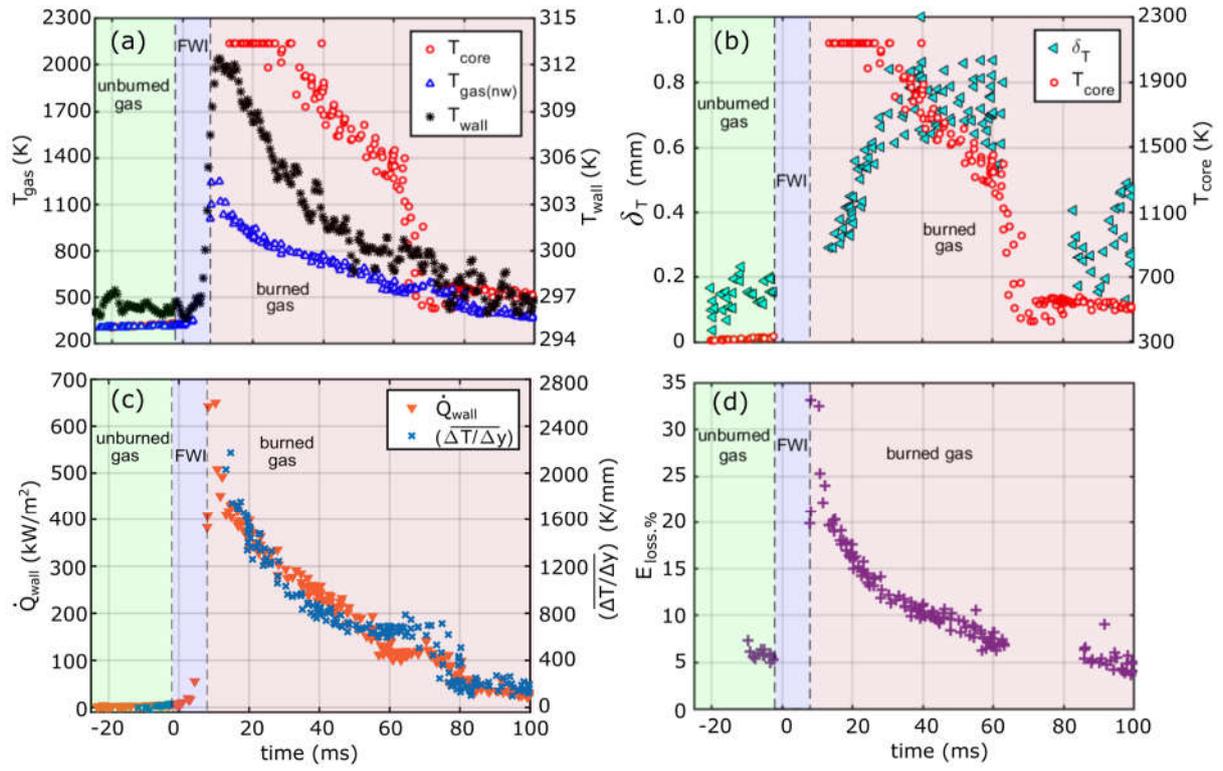

**Figure 5:** Time-history of relevant boundary layer and heat transfer quantities throughout the experiment duration.

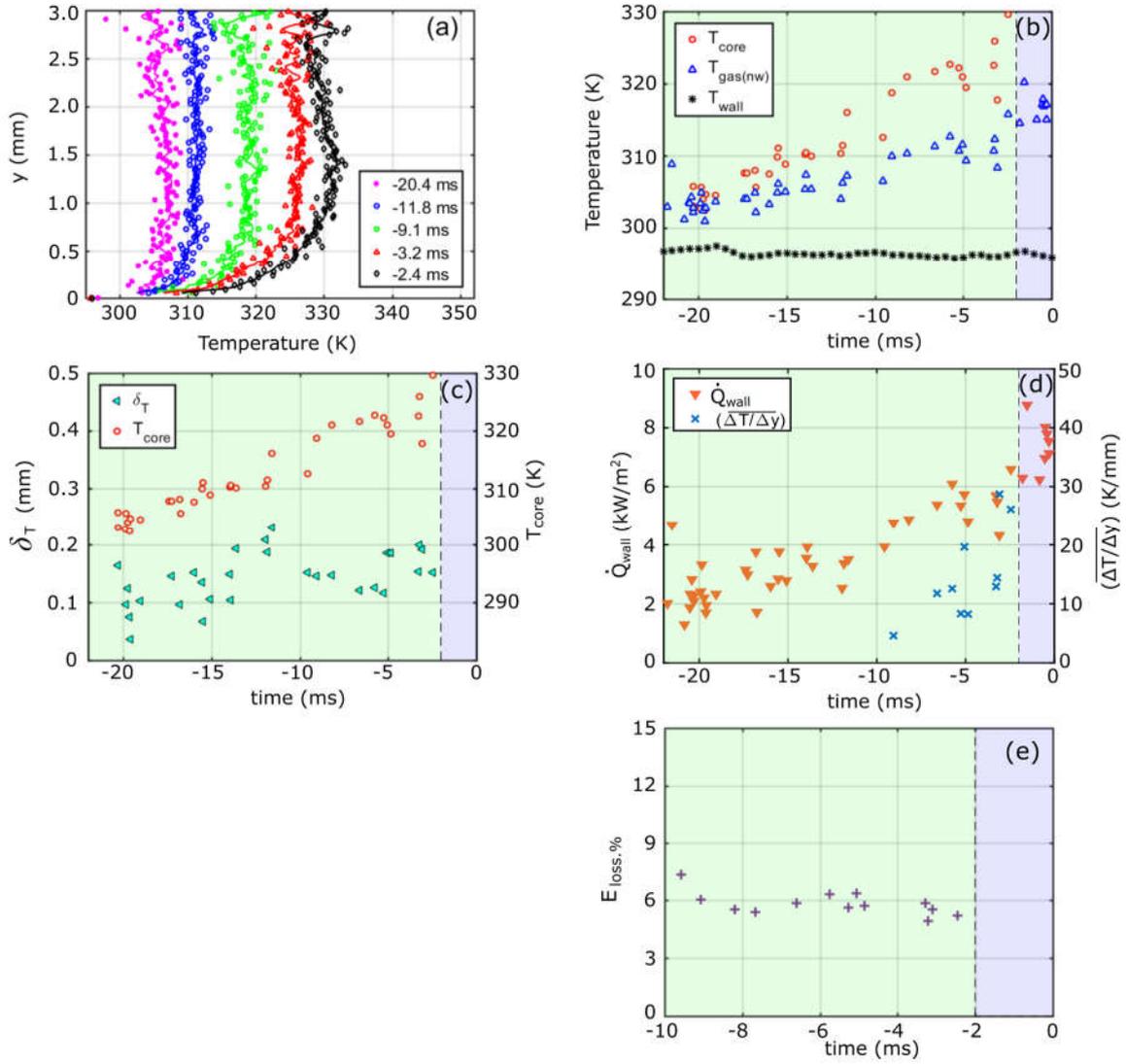

**Figure 6:** (a) selected 1D-temperature profiles during the unburned-gas regime. $T_{wall}$ is indicated at $y = 0$. (b) – (e) time-history of relevant quantities characterizing the boundary layer and wall heat loss. Lines plotted through data in (a) represent 5-pt (100 μm) moving average.

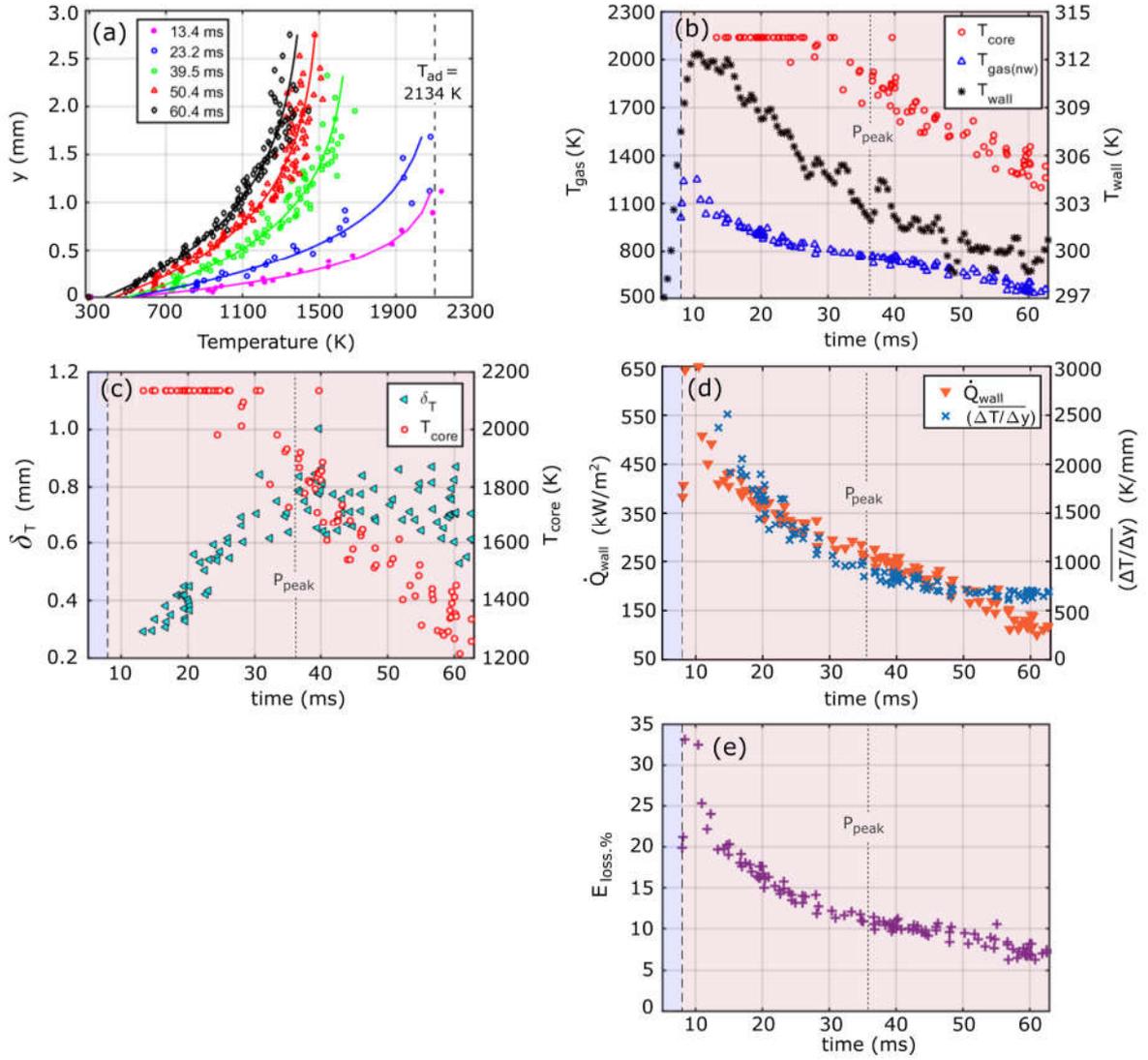

**Figure 7:** (a) selected 1D-temperature profiles during the post-FWI regime. $T_{wall}$ is indicated at $y = 0$. (b) – (e) time-history of relevant quantities characterizing the boundary layer and wall heat loss. Lines plotted through data in (a) represent data fits described by Eqn. (2).

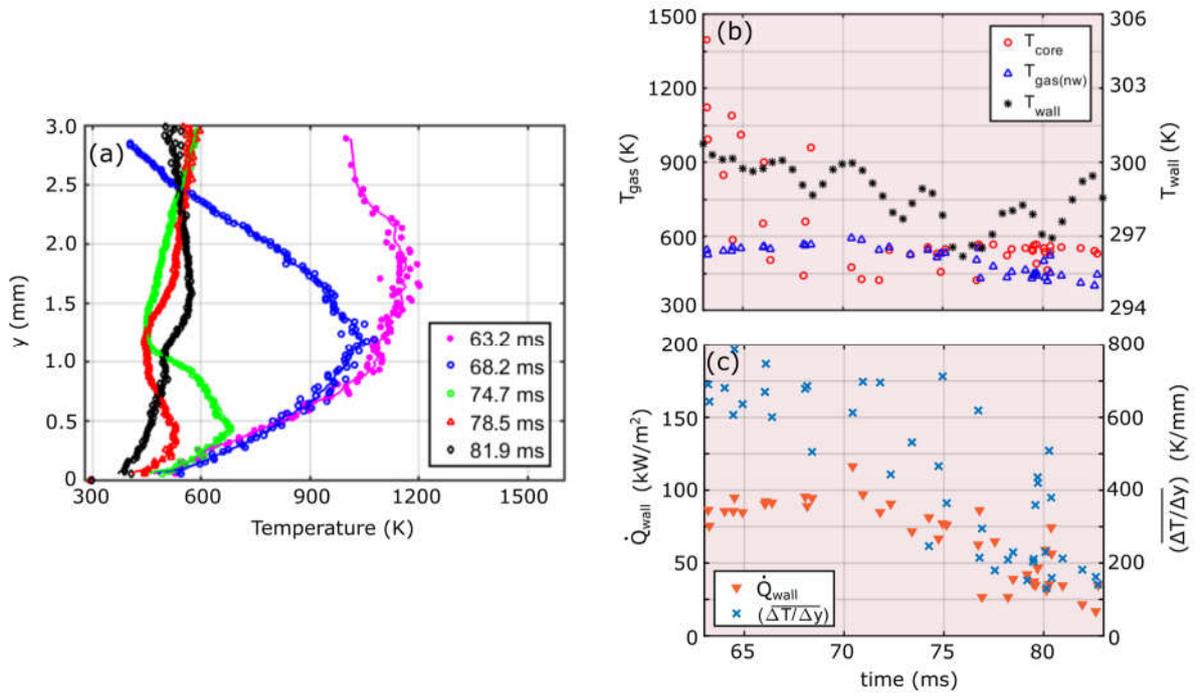

**Figure 8:** (a) selected 1D-temperature profiles during the outgassing regime. $T_{wall}$ is indicated at $y = 0$. (b) and (c) time-history of relevant quantities characterizing the boundary layer and wall heat loss. Lines plotted through data in (a) represent 5-pt (100 µm) moving average.

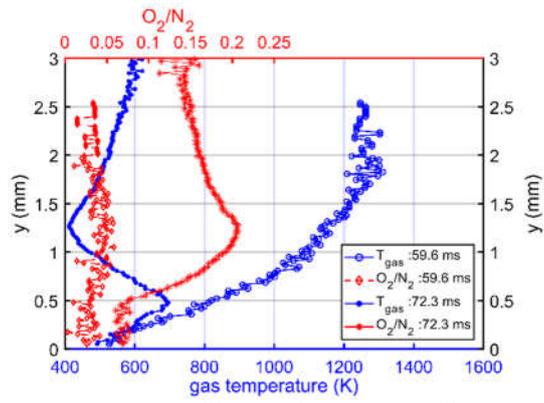

**Figure 9:** 1D gas temperatures profiles and corresponding 1D $O_2/N_2$ profiles for selected times before and during the outgassing regime.

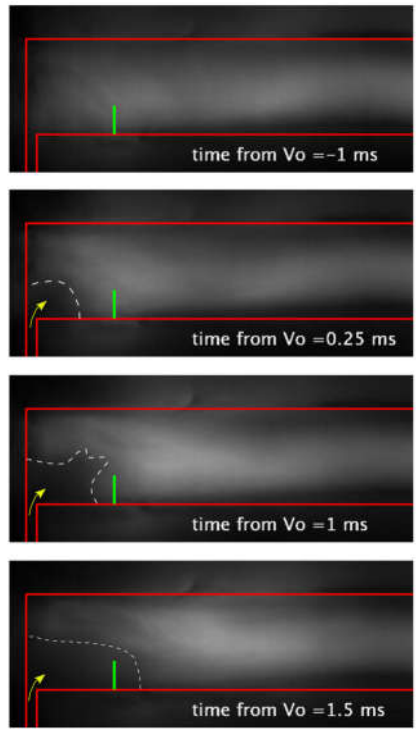

**Figure 10:** High-speed luminescence images describing the low intensity gas exiting the crevice during outgassing. Operating conditions: $C_2H_4$, 2.03 bar initial pressure, and 1.25 mm crevice spacing. Timing is reference to valve opening (Vo). White dash line depicts extent of lower intensity region exiting crevice region.

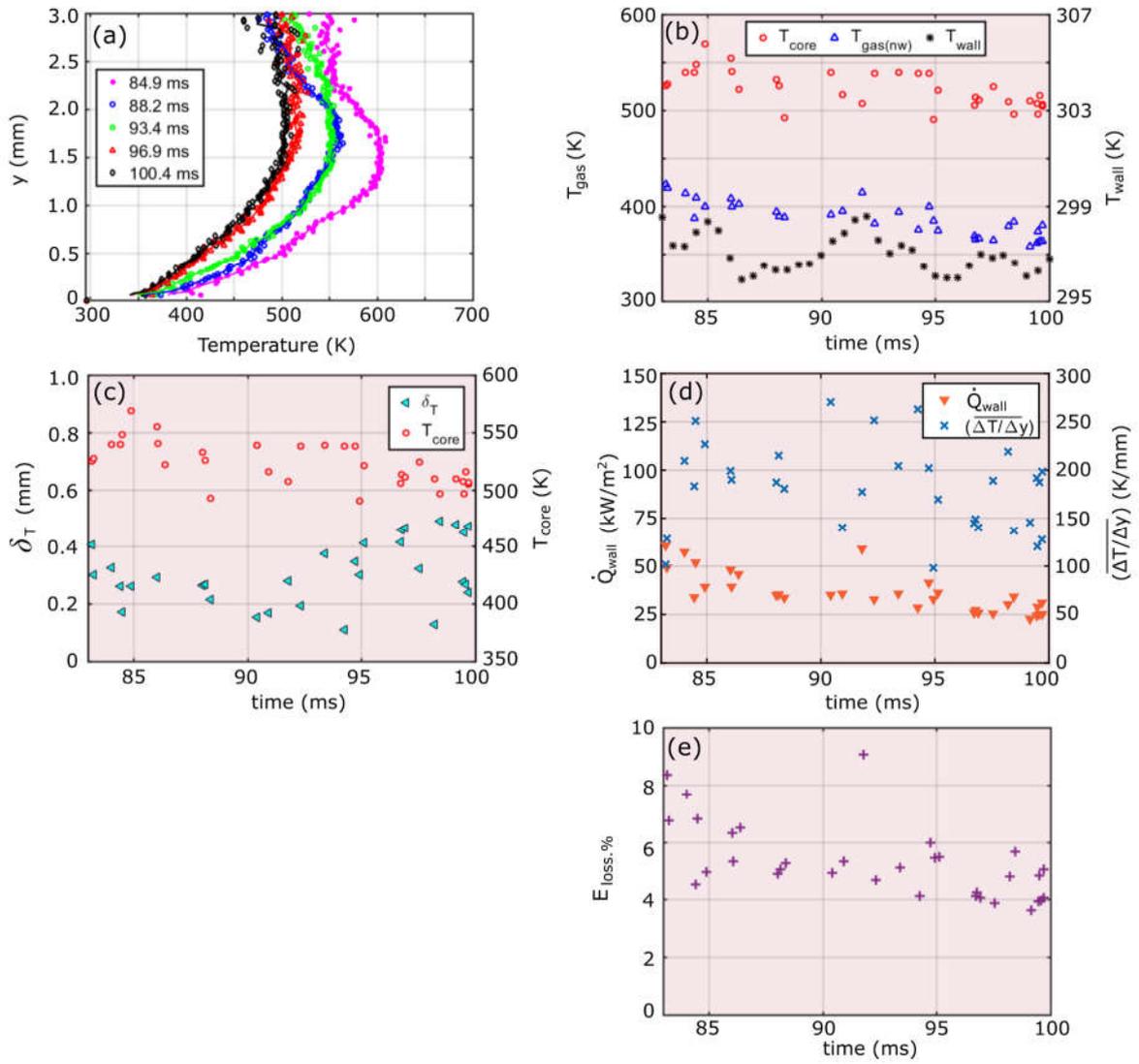

**Figure 11:** (a) selected 1D-temperature profiles during the end-exhaust regime. $T_{wall}$ is indicated at $y = 0$. (b) – (e) time-history of relevant quantities characterizing the boundary layer and wall heat loss. Lines plotted through data in (a) represent 5-pt (100 μm) moving average.

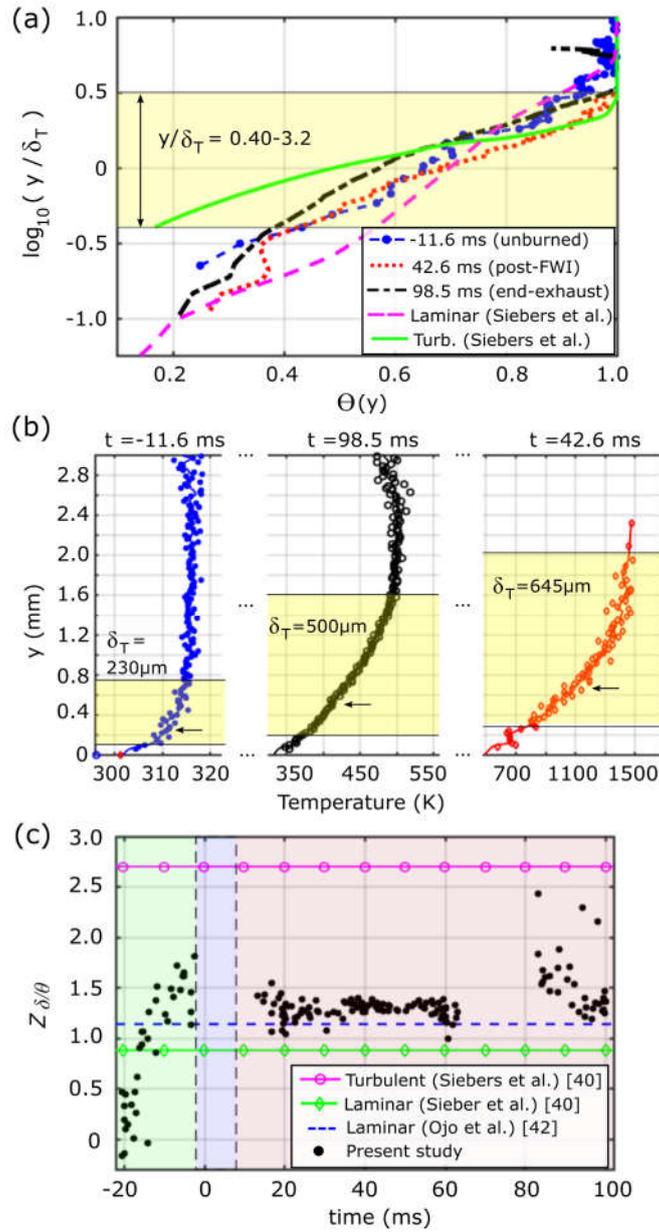

**Figure 12:** (a) normalized temperature profiles vs. wall normal distance. Steady-state laminar and turbulent cases are from [40]. Slope, $z_{\delta/\theta}$, extracted from $0.4 \leq y/\delta_T \leq 3.2$. (b) 1D-temperature profiles for selected FVC cases shown in (a). $0.4 \leq y/\delta_T \leq 3.2$ region is highlighted. (c) time-history of $z_{\delta/\theta}$ in comparison to canonical cases from [40] and unsteady laminar boundary layers from [42].